\documentclass[12pt]{emulateapj}

\begin{document}

%\title{Cycle variations of the acoustic solar radius estimated by seismic holography}
\title{Meridional Circulation During the Extended Solar Minimum: Another Component of the Torsional Oscillation?}


\author{I. Gonz{\'a}lez Hern{\'a}ndez, R. Howe, R. Komm and F. Hill}
%% \altaffilmark{1}}
\affil{ National Solar Observatory\footnote
	{Operated by the Association of Universities for Research in 
	Astronomy, Inc. under cooperative agreement with the National 
	Science Foundation.}, \\ 950 N. Cherry Ave., Tucson, AZ, USA}

%\author{C. Lindsey}
%\author{D. C. Braun}
%\affil{Northwest Research Associated, Boulder, CO, USA}

\email{irenegh@nso.edu}

\begin{abstract}

We show here a component of the meridional circulation developing at medium-high latitudes (40-50$^{\circ}$) before the new solar cycle starts. Like the torsional oscillation of the zonal flows, this extra circulation seems to precede the onset of magnetic activity at the solar surface and move slowly towards lower latitudes. However, the behavior of this component differs from that of the torsional oscillation regarding location and convergence towards the equator at the end of the cycle. The observation of this component before the magnetic regions appear at the solar surface has only been possible due to the prolonged solar minimum. The results could settle the discussion as to whether the extra component of the meridional circulation around the activity belts, which has been known for some time, is or is not an effect of material motions around the active regions.
\end{abstract}


\section{Introduction}
Meridional circulation has became an important player in the flux-transport solar-dynamo models \citep{miesch2005}. Until the development of local helioseismology methods, the observation of meridional circulation was limited to the surface layers by tracing magnetic elements  \citep{komm1993,snodgrass1996, nesme1997} or by using surface Doppler velocities \citep{hathaway1996}. \citet{giles1997} showed the first meridional circulation inferences under the solar surface by applying a particular local helioseismology technique, time distance. Although the full picture of the meridional circulation throughout the convection zone remains elusive, and recent studies suggest that very long series of data are required to infer the flows deep down \citep{braun2008}, local helioseismology  has been able to give detailed information on these flows in the subsurface layers. In particular, the variation with the solar cycle has been the focus of several studies \citep{haber2002,basu2003,gonzalez2006,zaatri2006,gonzalez2008}, which have shown that the overall amplitude of the meridional circulation is anticorrelated with the magnetic activity. This trend has been confirmed by tracing weak magnetic elements on the solar surface \citep{hathaway2010}.

\citet{chou2001} presented the first evidence of an additional subsurface component of the meridional circulation that varied with the solar cycle located around the center of the sunspot distribution. Applying the time-distance technique to observations from the Taiwan Oscillation Network (TON), they identified a residual of the meridional flow that correlated with the location of the surface activity for the declining phase of solar cycle 22 and the increasing phase of solar cycle 23. \citet{beck2002} confirmed the discovery using Michelson Doppler Imager (MDI) data from 1996 to 2001, and suggested this extra component could be a new component of the solar dynamo. In both cases, the authors conclude that this new component could be related to the well known phenomena of material moving inward/outward from active regions, although they noticed that the inferences suggested a larger than expected depth range.
\citet{gizon2003} tested the possibility of the extra circulation being related to the inflows around large active regions recorded by local helioseismology analysis \citep{zhao2001,haber2004,komm2004,braun2004}. He analyzed two Carrington rotations eliminating the contribution from areas with surface magnetic activity and found that the extra component decreased.

The continuous operations of the Global Oscillation Network Group (GONG) throughout the full solar cycle 23 has given us the unprecedented opportunity to study the subsurface meridional circulation without interruption.  Using these data, \citet{gonzalez2008} revealed that even after agressively masking surface magnetic activity over a 5.5-year period (2001-2006), the extra component of the meridional circulation remained present, although attenuated, under the solar surface. Here we extend this analysis to the prolonged solar minimum and find that the formation of the bumps at medium latitudes precedes the onset of the activity, confirming the nature of this  phenomenon as independent of the surface manifestation of active regions.

\section{Data Analysis}

%******************************
\begin{figure*}
\epsscale{.95}
\plotone{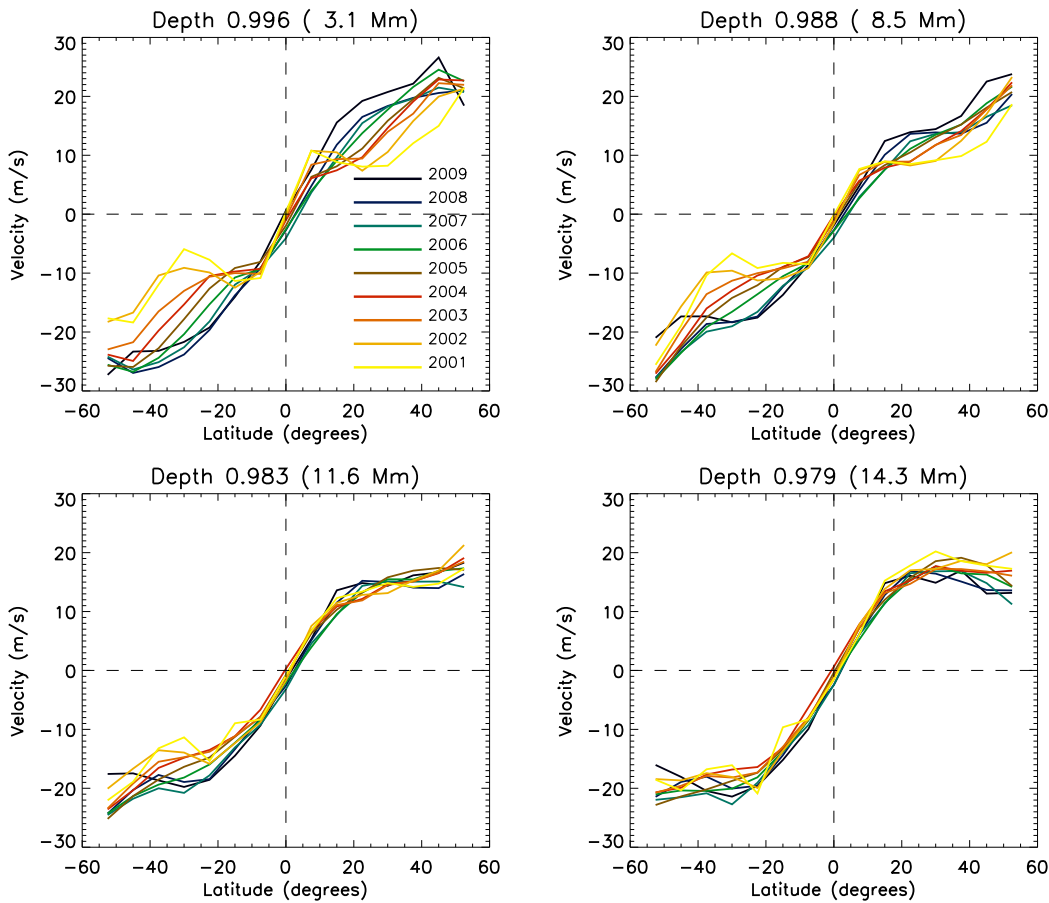}
\caption{Yearly averages of the meridional flow obtained by ring-diagram analysis of continuous GONG data at four different depths. The variation with the solar cycle clearly observed at the superficial layers is less pronounced at deeper layers. The extra circulation (bumps) is also clearly visible in the shallow layers. \label{fig:merid_yearly}}
\end{figure*}
%******************************
We extend the study  described by \citet{gonzalez2008} to years 2007, 2008 and 2009 to include the recent prolonged solar minimum. In summary, we apply standard ring-diagram analysis \citep{hill1988}, a local helioseismology technique, to GONG high-resolution Dopplergrams to infer the meridional flow from the solar surface to a depth of approximately 16\,Mm.

The ring diagrams  method uses medium/high-degree waves propagating in localized areas of the Sun to obtain an averaged horizontal-velocity vector for that particular region. The input data used for the analysis are high resolution Dopplergrams. In the typical  analysis, a particular area of the Sun (16$^{\circ}$ square, apodized to 15$^{\circ}$--diameter) is tracked and remapped,  and the analysis of the corresponding tridimensional power spectrum of solar oscillations renders an average horizontal velocity vector for the area at different depths \citep{corbard2003}. By analyzing a mosaic of these patches, it is possible to infer a three-dimensional velocity map in the depth range where the waves propagate. Typical ring-diagram analysis uses 1664-minute series of data  with a resolution of about 1.5 Mm per pixel at the center of the disk. We obtain horizontal velocity maps from disk center to approximately 52.5$^{\circ}$ in each direction with patches centered every 7.5$^{\circ}$ \citep{haber2002}. To study the meridional circulation we isolate the $v_{y}$ component of the calculated flows and use patches centered at no more than 22.5$^{\circ}$ heliocentric longitude to minimize projection effects.

Figure~\ref{fig:merid_yearly} presents the yearly-averaged meridional flows calculated for the period of August 2001 to April 2009. The results show the well known variation of the overall amplitude of the meridional circulation with the solar cycle: smaller amplitudes in 2001-2002 (maximum activity) and larger amplitudes towards 2008-2009 (minimum activity). The variation with the solar cycle is less pronounced at deeper depths, but it is still visible at high latitudes. \citet{gizon2008} present a model to account for independent observations of the meridional flow at the surface and at a depth of 60\,Mm. In such model, the  meridional flow at these two depths are anticorrelated. Although our analysis only returns information over the upper 15\,Mm, the depth dependence of the results is such that it would be consistent with their findings, with a reversal in the time varying component at depths below our accessible range.

\begin{figure}
\epsscale{1.}
\plotone{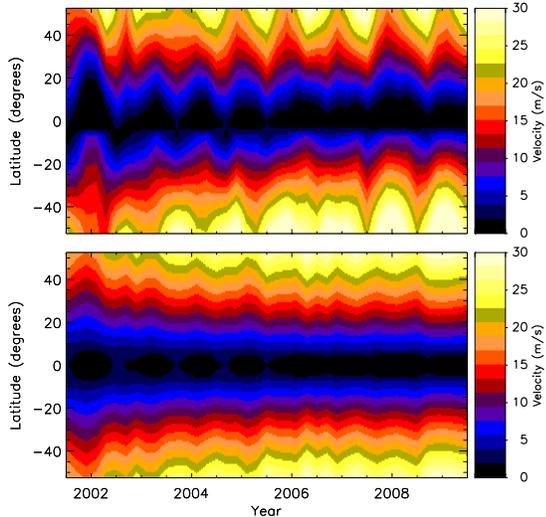}
\caption{Temporal variation of the fitted polynomial to the meridional circulation observations at a depth of 5.8\,Mm. A symmetrical plot averaging both hemispheres is shown in the bottom panel. Positive velocities are taken towards each respective pole.\label{fig:merid_butter}}
\end{figure}

Fitted low order polynomials to 60-day averages of the meridional flow at a depth of 5.8\,Mm are shown in Figure~\ref{fig:merid_butter}. Positive velocities are taken from the equator towards each pole.  Differences between the northern and southern hemisphere can be seen in the top panel. It is important to notice that local-helioseismology inferred flows at high latitudes have been shown to be affected by the periodic variation of the solar B$_{0}$-angle \citep{gonzalez2006, zaatri2006, beckers2007}. Although we do not yet have a full understanding of the effect, it is suspected to be related to the foreshortening in the data associated with the variation in the solar B$_{0}$-angle, the inclination of the solar rotation axis towards Earth. This periodic variation is clearly visible in the top panel of Figure~\ref{fig:merid_butter}. This systematic effect definitely affects our results; however, it should be consistent throught the solar cycle and hence it can not be responsible for the observed amplitude variation. An asymmetry between the North and South hemisphere is present in the results, in particular the fact that at the shallow layers the flows deviate from zero at the equator. The neutral line seems to be displaced towards the South during this period, which coincides with a more active period of the southern hemisphere (see Fig~\ref{fig:butter}).
 
The bottom panel of Figure~\ref{fig:merid_butter} shows a symmetrical plot of the flows, by averaging the North and South hemisphere. This reduces the systematic effect of the B$_{0}$-angle, without correcting it. The amplitude variation of the meridional flow, anticorrelated with the solar cycle, is more clearly visible here.

The extra circulation, in the form of {\it bumps}, superimposed on the smooth pattern of the meridional circulation, can also be seen in Figure~\ref{fig:merid_yearly} to a depth of $\sim$\,11\,Mm. For the earlier years, the bumps are centered in latitudes close to $\sim$\,30$^{\circ}$, towards the end of 2006, they have almost disappeared, only to appear again at $>$\,40$^{\circ}$ at the beginning of 2007.

\section{A Meridional Component of the Torsional Oscillation?}

%******************************
%******************************

\begin{figure}
\epsscale{1.}
\plotone{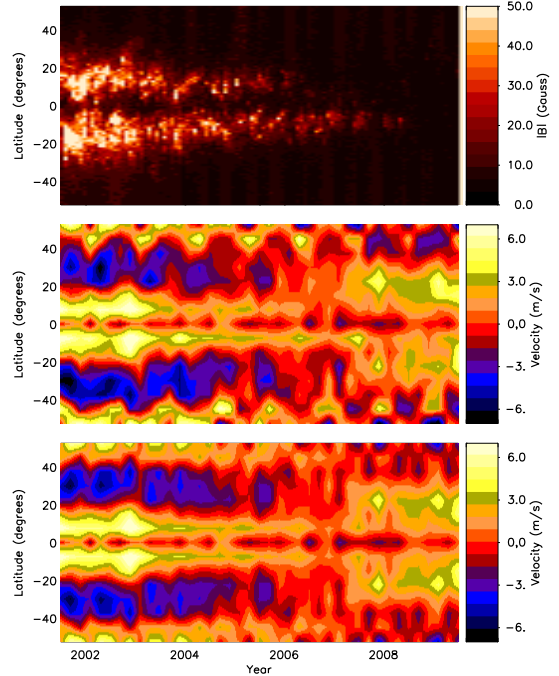}
\caption{Temporal variation of the meridional circulation residuals at a depth of 5.8\,Mm (central panel). Positive velocities are directed towards the poles. A symmetrical plot averaging both hemispheres is shown in the bottom panel.  The top panel shows the location and magnetic strength of the activity during the same period (calculated from MDI synoptic magnetograms).\label{fig:butter}}
\end{figure}

The residuals from a low order polynomial fitting to the inferred meridional circulation at a depth of 5.8\,Mm are shown in Figure~\ref{fig:butter}. The top panel is a proxy to the corresponding surface activity, calculated using a longitudinal average of MDI synoptic magnetograms over the studied period. The central panel presents the residuals averaged over 60 days and using a bilinear interpolation (IDL, REBIN function) for imaging the results. The bottom panel show a symmetric version, averaging North and South hemispheres. Positive velocities in both cases are taken towards the poles. 
The variation of this extra component of the meridional flows with the solar cycle resembles, to some extent, the behavior of the so-called solar torsional oscillation of the zonal flows during this extended solar minimum \citep{howe2009}. It appears at medium latitudes ($\sim$ 40-50$^{\circ}$) around 2007 and seems to be moving towards lower latitudes with time. Is this a meridional component of the torsional oscillation?

%******************************
%******************************
The appearance of the bumps at high latitudes before the magnetic activity surfaces happens at a similar time as the onset of the torsional oscillation,  with the center between faster and slower bands in 2007 located around 40$^{\circ}$ latitude. However, the amplitude of these residuals seem consistently smaller than those from the torsional oscillation, although marginally. Also, unlike the torsional oscillation, the extra component of the meridional circulation does not seem to converge towards  the equator at the end of the solar cycle.The bumps are also less clear below a certain depth and seem to disappear in the range attained with our analysis, although this maybe an artifact of the smoothing in the inversion technique. The results from global helioseismology analysis show that the torsional oscillation of the zonal flows remain present throughout the convection zone \citep{vorontsov2002,howe2005}; ring-diagram analysis shows little variation of the zonal flow amplitude in the outer 15\,Mm. The center of mass of the perturbance seems to be stable above 20 degrees latitude. Even with our limited spatial resolution, if the bumps were located directly under the surface activity, the north and southern residual branches would be much closer to the equator towards the end of the cycle. 

\begin{figure}
\epsscale{1.}
\plotone{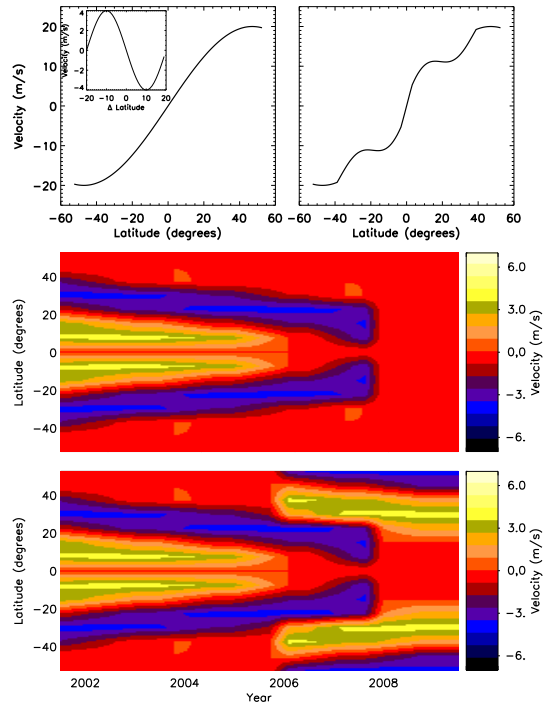}
\caption{Illustration of the extra meridional circulation. Top panels show the assumed meridional circulation, the model bumps and the result when adding them both at about 20$^{\circ}$ latitude. The central panel presents the residuals, with the ring-diagram resolution, when moving the bumps acording to the location of surface activity. The bottom panel includes the developing branch at higher latitudes. These two simple experiments do not agree with the observations, at least close to the equator.\label{fig:model}}
\end{figure}

Figure~\ref{fig:model} represents an illustration  of the meridional flow (top panel, A) with superimposed bumps, inwards flows towards the center of activity (top panel, B). We move the bumps with time following the latitude location of surface activity, keeping the amplitude of bumps constant, and then subsample the results with the same resolution that we obtain from the ring-diagram analysis. To image the results, we use the same bilinear interpolation function (IDL, REBIN). The central panel shows what happens when the activity is only present at the surface. It can be seen that the residuals continue their convergence towards the equator after 2006, when they disappear from the real observations. To check if this was an effect of the new components at high latitudes, we repeat the experiment now including extra bumps centered at 50$^{\circ}$ in 2006 and moving slowly towards the equator. Neither of these two simple experiments agree with the observations. In the real data, the lower-latitude  bumps are suppressed from 2006 to 2008, when surface activity is still present.  However, a simple relation is not expected because, as found in previous work \citep{gizon2003,gonzalez2008}, removing the areas with surface activity changes the meridional circulation. Hence, there is also an expected contribution to the flows from active regions. Also, the coupling between zonal and meridional flows through the Coriolis force is latitude dependent.

Local helioseismology inferences of meridional circulation during  the previous minimum do not show clear evidence of this medium-latitude formation of extra circulation before the activity appears at the surface \citep{chou2001,beck2002}. Unfortunately, the availability of high resolution helioseismic data for local helioseismology methods dates back only to 1994, with the TON instrument, and to 1996 with MDI. This, combined with the short span between the end of solar cycle 22 and the beginning of solar cycle 23 and the long term averages needed to infer meridional circulation makes the isolation of the medium latitude branch difficult for that period. Thanks to this extended solar minimum we have been able to clearly observe the phenomenon before the onset of the new cycle.

%******************************
%******************************
\section{Discussion and Conclusions}

Our results confirm the solar cycle variation of the overall amplitude of the meridional circulation, minimum amplitude at maximum activity and vice versa. The trend is clearly there during the decline of solar cycle 23 and has continued during the current prolonged solar minimum. No clear sign of trend reversal is found in the studied period.

We also present, for the first time, a conclusive proof that the extra circulation, or bumps, of the meridional circulation under the solar surface is independent of the presence of surface magnetic activity.  \citet{spruit2003} presented a model that explains the torsional oscillation as a geostrophic flow due to the lower subsurface temperature in active regions. The model predicts the appearance of flows from the edges towards the center of the main latitude of magnetic activity, a meridional version of the torsional oscillation with a maximum amplitude of $\sim$6\,m/s at the surface. The model also predicts a rapid decline of these oscillations with depth, which would disappear below 30\,Mm. Our results are consistent in amplitude with this model. However, we show a more rapid attenuation of the inflows with depth, which disappear around 10-14\,Mm and we are observing these flows at medium-high latitudes in the absence of surface activity.

The extra component at medium-high latitudes starts to develop in 2007 for the new cycle, and continues there for the rest of the extended minimum. With our limited resolution and data span, it is difficult to conclude if there is a variation in the latitude of the extra component for the last two years, although it seems to be migrating slowly towards lower latitudes. Applying ring-diagrams method to the current helioseismic data it is not possible to reach latitudes higher than $\sim$50$^{\circ}$, hence we can not distinguish the possible formation of this medium-latitude new branch earlier on at higher latitudes. The upcoming launch of the Solar Dynamics Observatory (SDO) with the higher-resolution Helioseismic and Magnetic Imager (HMI) instrument on board will help to study meridional circulation above our present limits.

The amplitude of the meridional residuals seems consistently smaller than those from the torsional oscillation. A full, comprenhensive model that include both the zonal and the meridional components of the residual circulation is necessary to understand the dynamic implication of this cycle varying component.

\section*{Acknowledgments}

The authors want to thank S. Kholikov and J. Leibacher for useful discussions. This work utilized data obtained by the Global Oscillation Network Group (GONG) program, managed by the National Solar Observatory, which is operated by AURA, Inc. under a cooperative agreement with the National Science Foundation. The data were acquired by instruments operated by the Big Bear Solar Observatory, High Altitude Observatory, Learmonth Solar Observatory, Udaipur Solar Observatory, Instituto de Astrof{\'{\i}}sica de Canarias, and Cerro Tololo Interamerican Observatory. Magnetograms from the SOI/MDI on SOHO have also been used for the analysis. SOHO is a project of international collaboration between ESA and NASA. This work has been supported by the NASA Living with a Star - Targeted Research and Technology program - and the Stellar Astrophysics branch of the National Science Foundation.

\end{document}